\documentclass[aps,prb,onecolumn,floats,superscriptaddress]{revtex4}
\usepackage[pdftex]{graphicx} % Include figure files
\usepackage{mathtools}
\usepackage{txfonts}
\usepackage{float}
\usepackage[linkcolor=black,anchorcolor=black,citecolor=black,hypertexnames=false]{hyperref}
\usepackage{multibib}
\usepackage{amssymb}
\usepackage{multirow}
\renewcommand\figurename{\textbf{Fig.}}
\newcommand{\bra}[1]{\ensuremath{\langle#1|}}
\newcommand{\ket}[1]{\ensuremath{|#1\rangle}}
\begin{document}
\large
\title{Intertwined dipolar and multipolar order in the triangular-lattice magnet TmMgGaO$_4$}

\author{Yao Shen}
\affiliation{State Key Laboratory of Surface Physics and Department of Physics, Fudan University, Shanghai 200433, China}

\author{Changle Liu}
\affiliation{State Key Laboratory of Surface Physics and Department of Physics, Fudan University, Shanghai 200433, China}

\author{Yayuan Qin}
\affiliation{State Key Laboratory of Surface Physics and Department of Physics, Fudan University, Shanghai 200433, China}

\author{Shoudong Shen}
\affiliation{State Key Laboratory of Surface Physics and Department of Physics, Fudan University, Shanghai 200433, China}

\author{Yao-Dong Li}
\affiliation{State Key Laboratory of Surface Physics and Department of Physics, Fudan University, Shanghai 200433, China}
\affiliation{Department of Physics, University of California Santa Barbara, Santa Barbara, California 93106, USA}

\author{Robert Bewley}
\affiliation{ISIS Facility, Rutherford Appleton Laboratory, STFC, Chilton, Didcot, Oxon OX11 0QX, United Kingdom}

\author{Astrid Schneidewind}
\affiliation{J$\ddot{u}$lich Centre for Neutron Science (JCNS) at Heinz Maier-Leibnitz Zentrum (MLZ), Forschungszentrum J$\ddot{u}$lich GmbH, Lichtenbergstr. 1, 85748 Garching, Germany}

\author{Gang Chen$^\ast$}
\affiliation{State Key Laboratory of Surface Physics and Department of Physics, Fudan University, Shanghai 200433, China}
\affiliation{Department of Physics and Center of Theoretical and Computational Physics, The University of Hong Kong, Pokfulam Road, Hong Kong, China}
\affiliation{Center for Field Theory and Particle Physics, Fudan University, Shanghai, 200433, China}
\affiliation{Collaborative Innovation Center of Advanced Microstructures, Nanjing University, Nanjing, 210093, China}

\author{Jun Zhao$^\ast$}
\affiliation{State Key Laboratory of Surface Physics and Department of Physics, Fudan University, Shanghai 200433, China}
\affiliation{Collaborative Innovation Center of Advanced Microstructures, Nanjing University, Nanjing, 210093, China}

\maketitle

\textbf{A phase transition is often accompanied by the appearance of an order parameter and symmetry breaking. Certain magnetic materials exhibit exotic hidden-order phases, in which the order parameters are not directly accessible to conventional magnetic measurements. Thus, experimental identification and theoretical understanding of a hidden order are difficult. Here we combine neutron scattering and thermodynamic probes to study the newly discovered rare-earth triangular-lattice magnet TmMgGaO$_4$. Clear magnetic Bragg peaks at K points are observed in the elastic neutron diffraction measurements. More interesting, however, is the observation of sharp and highly dispersive spin excitations that cannot be explained by a magnetic dipolar order, but instead is the direct consequence of the underlying multipolar order that is ``hidden" in the neutron diffraction experiments. We demonstrate that the observed unusual spin correlations and thermodynamics can be accurately described by a transverse field Ising model on the triangular lattice with an intertwined dipolar and ferro-multipolar order.}

Interactions between various microscopic degrees of freedom with similar energy scales can induce strong competition and frustration, leading to exotic phenomena. The f-electron materials show strong spin-orbit coupling (SOC), and therefore the spin and orbital degrees of freedom should be described by the total angular momentum $J$. The crystalline electric field (CEF) further splits the total angular momentum $J$, and the low-lying crystal field states can form rather complex ground states, including spin liquids~\cite{YMGO_YL1,YMGO_YL2,YMGO_YS1,YMGO_Martin,YMGO_YS2,NaYbS,112}, spin ices~\cite{Balents,Gingras,HoTiO}, and hidden ordered phases~\cite{MultiRev,URuSiRev,CeBRev}. In most circumstances, the hidden-order phase transition is signaled by the change of bulk properties such as the magnetic susceptibility and heat capacity~\cite{URuSi_HC}; but unveiling its microscopic nature is difficult because the hidden-order parameter cannot be directly disclosed by microscopic probes such as neutron diffraction or muon spin rotation/relaxation~\cite{MultiRev}. Efforts have been made through the study of the collective excitations associated with the hidden orders~\cite{CeBNS0,UO2,URuSiNS1,URuSiNS2,CeBNS}. For example, in the canonical hidden-order material URu$_2$Si$_2$, the antiferromagnetic spin excitations appear at both the commensurate and incommensurate wavevectors, and exhibit spin gaps in the hidden order phase~\cite{URuSiNS2}. In the case of CeB$_6$, complex ferromagnetic and antiferromagnetic spin excitations were observed along with a spin exciton~\cite{CeBNS}. However, whether or how these collective modes drive the hidden order phase transition remains a matter of debate.

The recent discovered rare-earth magnet TmMgGaO$_4$ may provide a new opportunity to examine the exotic ordering phenomenon of f-electrons~\cite{TMGO_Cava}. This material has the same crystal structure as the spin liquid candidate YbMgGaO$_4$ which crystallizes in the $R\bar{3}m$ space group with a quasi-two-dimensional triangular lattice~\cite{YMGO_YL1}. In TmMgGaO$_4$, the Tm$^{3+}$ ion possesses an electron configuration 4$f^{12}$, in which the orbital and spin angular momentum ($L$ = 5, $S$ = 1) are entangled into the total angular momentum $J$ = 6 due to the strong SOC. The corresponding 13 states are further split under the $D_{3d}$ CEF. It was suggested that the relevant low energy degrees of freedom in Tm$^{3+}$ ion is a pair of nearly degenerate singlets that are well separated from other CEF levels, resulting in a quasi-doublet state~\cite{TMGO_YL}. This indicates the magnetic interaction of TmMgGaO$_4$ should be different from that of YbMgGaO$_4$ where Yb$^{3+}$ ions are Kramers ions~\cite{Yaodong}. The thermodynamic and magnetic susceptibility measurements in TmMgGaO$_4$ further revealed a phase transition at low temperature and suggested a conventional stripe-type dipolar magnetic order state with Ising interactions~\cite{TMGO_YL}, which has yet to be verified by microscopic measurements.

In this paper, we report neutron scattering and thermodynamic studies of the phase transitions and spin correlations of TmMgGaO$_4$ single crystals. Our data reveal that the longitudinal component of the effective spins, $S^z$, presents dipolar behavior and forms a three sublattice magnetic structure, leading to a magnetic Bragg peak at the K point in the neutron diffraction spectrum. The transverse components, $S^x$ and $S^y$, however, exhibit multipolar behavior and develop a ferro-multipolar order, which is mostly hidden in elastic neutron diffraction, but is revealed by the collective spin excitations in the inelastic channel. We show that the spin correlations and thermodynamics of TmMgGaO$_4$ can be accurately described by the transverse field Ising model with an intertwined dipolar and multipolar order on the triangular lattice.

\textbf{Results}

\begin{figure*}[t]
\includegraphics[width=0.8\textwidth]{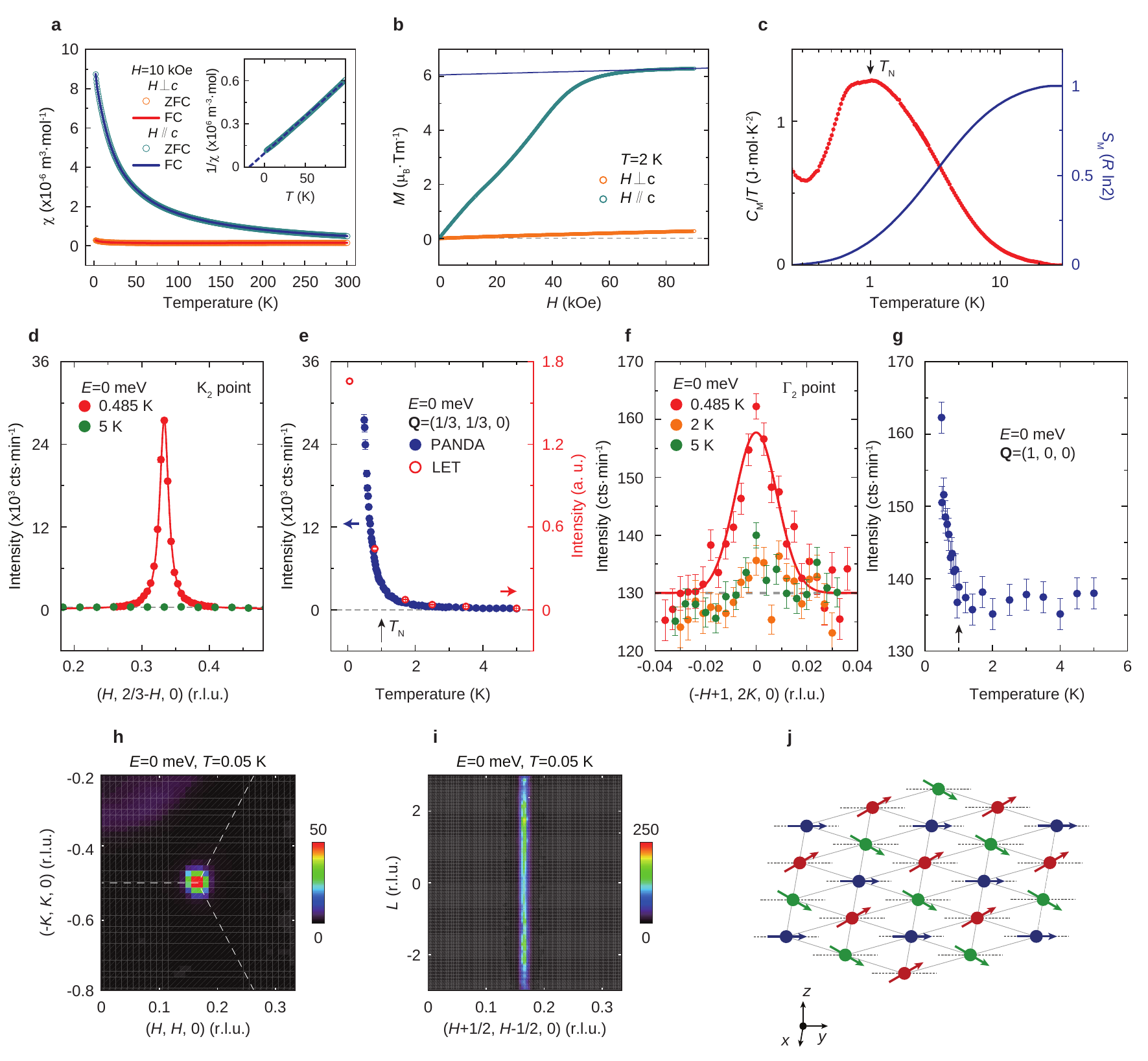}
\caption{\textbf{Thermodynamic property and neutron diffraction measurements of TmMgGaO$_4$ single crystals.} \textbf{a,} Temperature dependence of the magnetic susceptibility $\chi$ measured under ZFC and FC conditions with external fields of 10 kOe applied parallel and perpendicular to the $c$ axis. The inset shows the linear fitting of the inverse susceptibility. \textbf{b,} Field dependence of the magnetization at $T$ = 2 K. Linear fitting of the magnetization at high field indicates the Land\'{e}-$g$ factor of 12.11(5). \textbf{c,} Magnetic heat capacity and magnetic entropy measured under zero field. The phonon contribution is subtracted by comparing measurements of TmMgGaO$_4$ with the non-magnetic reference compound LuMgGaO$_4$. The magnetic entropy is obtained by integrating $C/T$ from 0.25 K. Indication of a Schottky anomaly is observed below 0.4 K, which is likely caused by the strong hyperfine interactions. \textbf{d,} $\textbf{Q}$-scans across the magnetic Bragg peak $\textbf{Q}$ = (1/3, 1/3, 0) along the transverse direction at the indicated temperatures. \textbf{e,} Temperature dependence of the fitted peak amplitudes of the Bragg peak at $\textbf{Q}$ = (1/3, 1/3, 0). \textbf{f,} $\textbf{Q}$-scans across the multipolar Bragg peak $\textbf{Q}$ = (1, 0, 0) along the transverse direction at indicated temperatures. \textbf{g,} Temperature dependence of the intensity of the $\textbf{Q}$ = (1, 0, 0) peak. The solid and dashed lines in \textbf{d}-\textbf{f} are guides to the eye. \textbf{h, } Momentum dependence of the magnetic Bragg peak at 0.05 K. The white dashed lines indicate the zone boundaries. \textbf{i,} $L$ dependence of the peak intensity at $\textbf{Q}$ = (2/3, -1/3, $L$). The color bars indicate scattering intensity in arbitrary unit in linear scale. \textbf{j,} Schematic of the three-sublattice magnetic structure of TmMgGaO$_4$. $S^y$ forms ferro-multipolar order along the $y$ direction (black dashed lines) and $S^z$ forms dipolar order (spin up-red, spin 0-blue, spin down-green). The red and green arrows are tilted from the $xy$ plane by $\sim32$ degrees. The data shown in \textbf{d}, \textbf{f} and \textbf{g} were measured on PANDA and the data in \textbf{h} and \textbf{i} were measured on LET. The wavevector \textbf{Q} is defined as $\textbf{Q}=H\textbf{a}^*+K\textbf{b}^*+L\textbf{c}^*$; r.l.u., reciprocal lattice unit; cts$\cdot$ min$^{-1}$, counts per minute; error bars, 1 s.d.
}
\label{fig1}
\end{figure*}

\textbf{Susceptibility, magnetization and heat capacity.} We first measure the magnetic susceptibility of our TmMgGaO$_4$ single crystals prepared by the floating zone technique (Method). With an external field applied along the $c$ axis, paramagnetic behavior is revealed at low temperatures with a Curie-Weiss temperature $-19.1$ K (Fig.~\ref{fig1}a). No clear anomaly or splitting of the zero-field-cooling (ZFC) and field-cooling (FC) data can be distinguished, suggesting the absence of phase transition above 2 K. When the magnetic field along the $c$ axis increases, the system is driven into a nearly polarized state marked by the saturated magnetization above 5 T (Fig.~\ref{fig1}b). The corresponding Land\'{e}-$g$ factor for effective $S$=1/2 state is around 12.11 which is close to the upper limit of $2Jg_J=14$. This indicates a dominant $J^z=\pm$6 components in the wave functions of the low-lying two singlets, leading to a strong Ising character, consistent with previous report~\cite{TMGO_YL}. In a sharp contrast, when the field is applied in the $ab$ plane, a much weaker magnetic response is observed (Fig.~\ref{fig1}a, b). The heat capacity measurement shows an anomaly at $\sim$ 1 K, indicating a phase transition (Fig.~\ref{fig1}c). The integral magnetic entropy below 30 K is close to $R\ln$2 which is expected for a system with $S_{\mathrm{eff}}$=1/2, in agreement with the quasi-doublet CEF ground state of the Tm$^{3+}$ ions in TmMgGaO$_4$.

\textbf{Magnetic Bragg peaks and spin wave dispersion.} To determine the microscopic ground state of TmMgGaO$_4$, we use elastic neutron diffraction to study the magnetic ordering properties. Clear magnetic Bragg peaks are uncovered at K points at low temperatures (Fig.~\ref{fig1}d, 1h), suggesting the presence of a magnetic dipolar order. For the scans along the $L$ direction at the K point, we observe a rod of scattering (Fig.~\ref{fig1}i), demonstrating the quasi-two-dimensional nature of the magnetic order. The peak intensity drops gradually on warming and shows a transition at $\sim$ 1 K, consistent with the heat capacity measurements  (Fig.~\ref{fig1}e).

\begin{figure*}[bth]
\includegraphics[width=\textwidth]{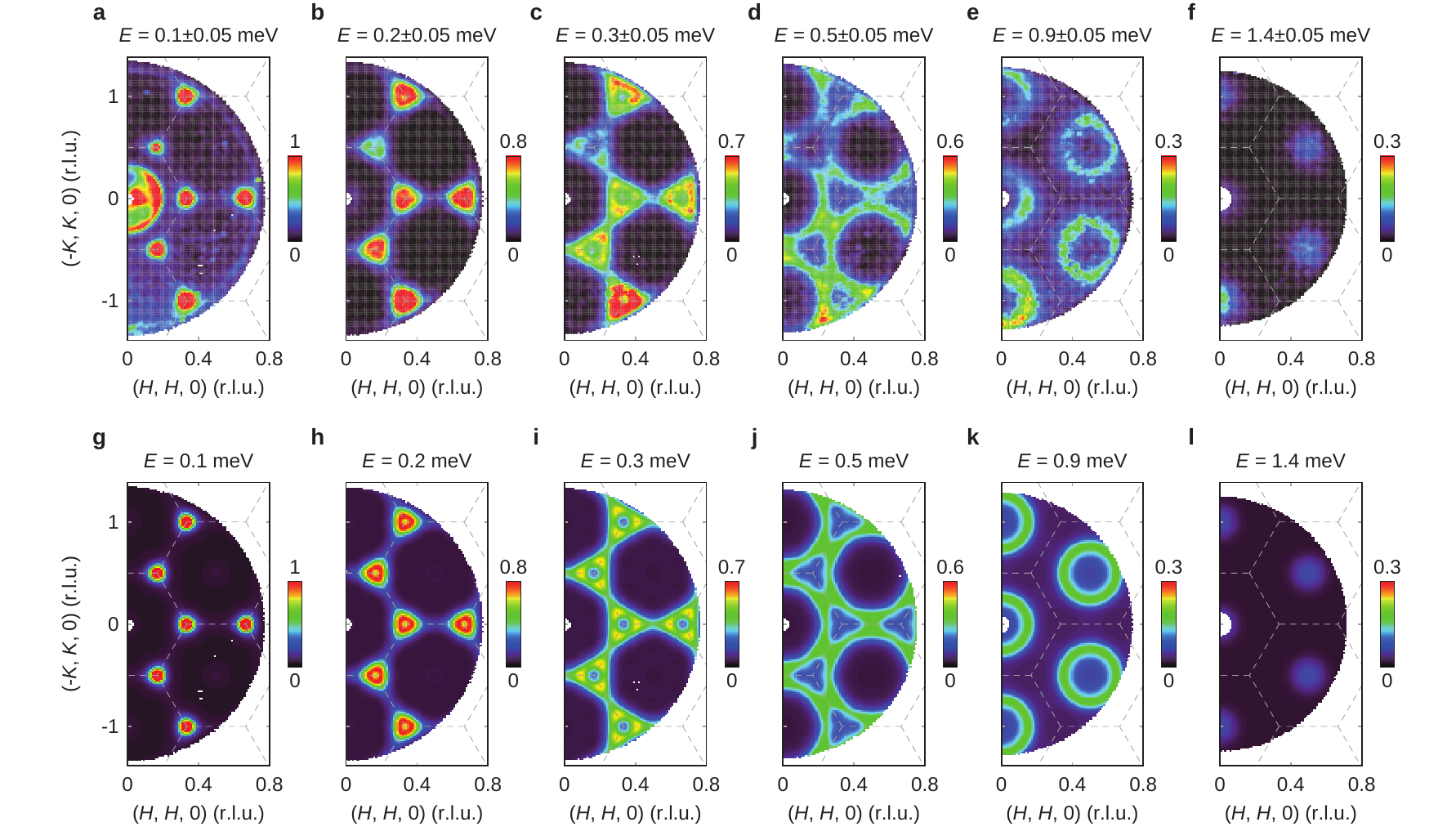}
\caption{\textbf{Measured and calculated momentum dependence of the spin excitations in TmMgGaO$_4$ at the indicated energies and ${T = 0.05}$ K.} \textbf{a-f,} Raw contour plots of the constant energy images at ${T = 0.05}$ K. The signals near \textbf{Q} = (0, 0, 0) in \textbf{a} are due to the elastic contamination from the sample environment close to the direct beam. \textbf{g-l,} Calculated spin excitations using the model specified in the text. The dashed lines indicate the zone boundaries. The measurements were performed on LET spectrometer with ${E_\textrm{i} = 4.8}$ meV. The color bars indicate scattering intensity in arbitrary unit in linear scale.
}
\label{fig2}
\end{figure*}

\begin{figure*}[bth]
\includegraphics[width=0.8\textwidth]{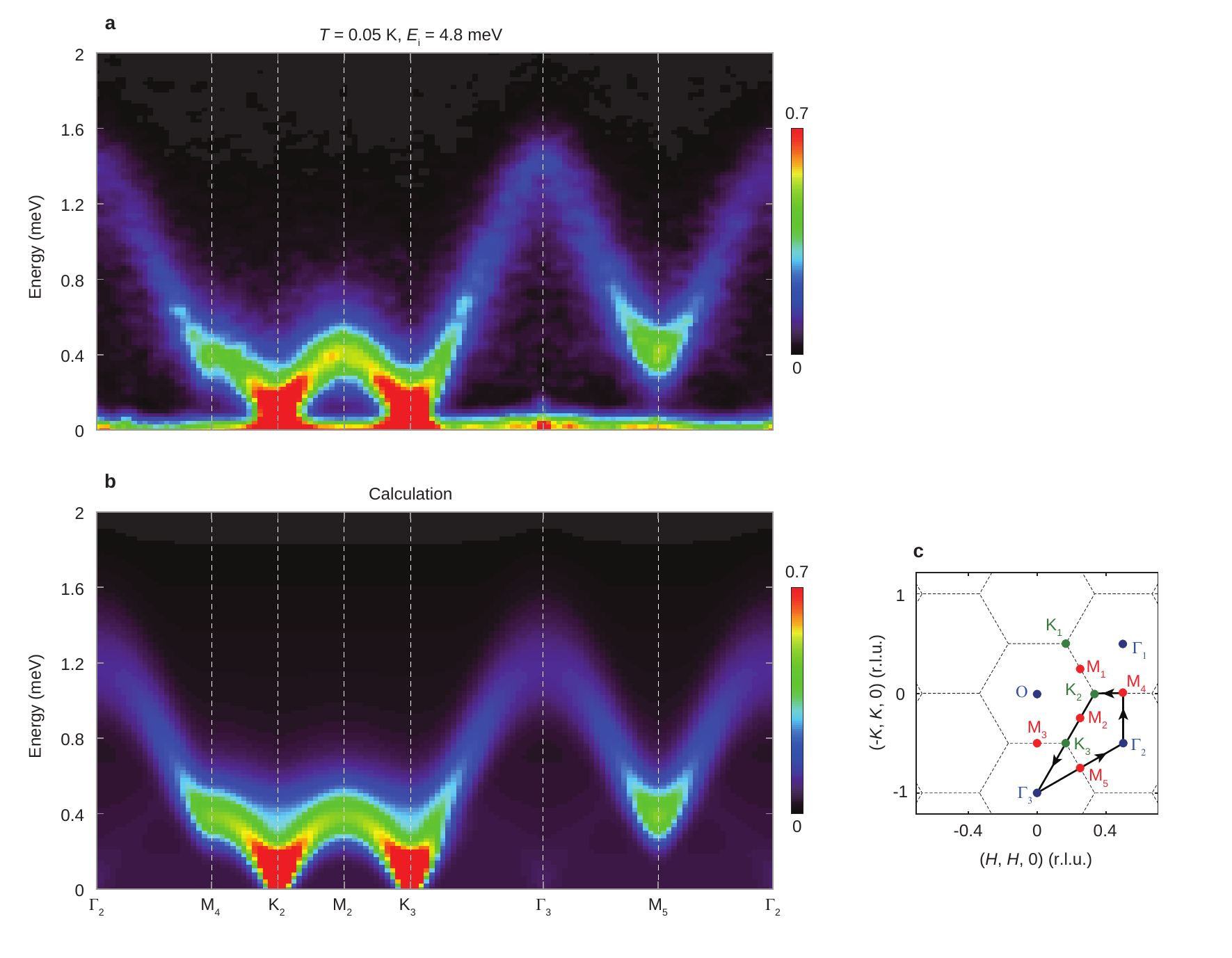}
\caption{\textbf{Measured and calculated spin wave dispersions in TmMgGaO$_4$ at ${T = 0.05}$ K.} \textbf{a,} Intensity of the spin-excitation spectra along the high-symmetry momentum directions as illustrated by the black solid lines in \textbf{c}. \textbf{b,} The simulated spin excitation dispersion using the model specified in the text. The color bars indicate scattering intensity in arbitrary unit in linear scale. \textbf{c,} Sketch of the reciprocal space. Black dashed lines indicate the Brillouin zone boundaries.
}
\label{fig3}
\end{figure*}

\begin{figure}[bth]
\includegraphics[width=0.45\textwidth]{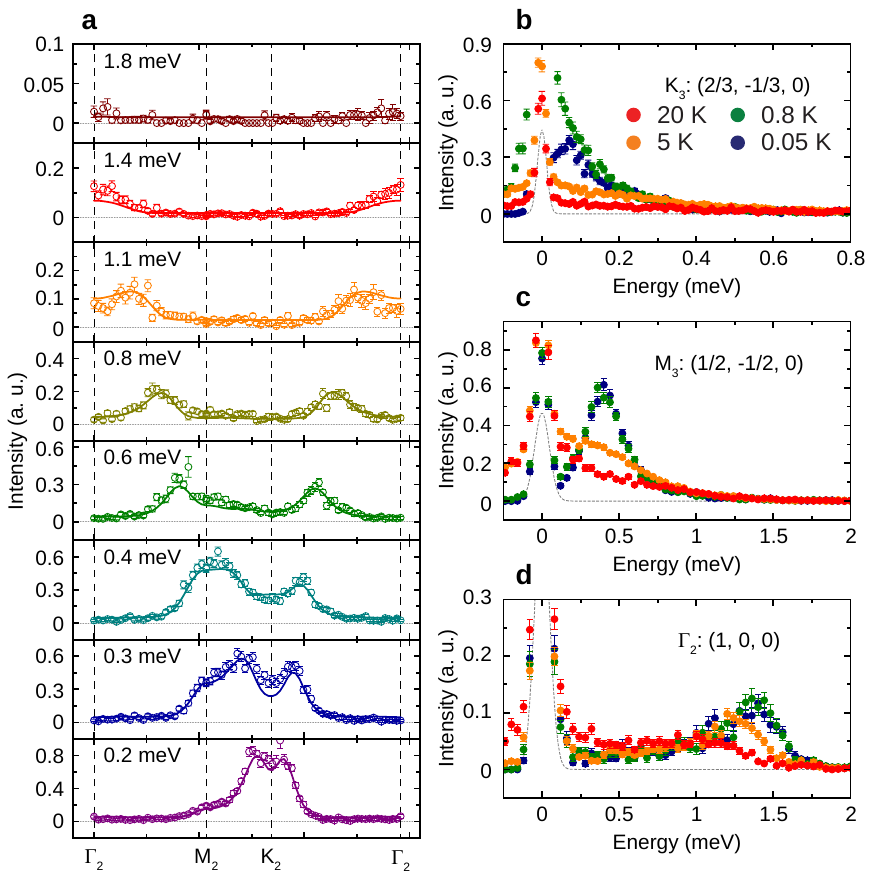}
\caption{\textbf{Constant energy cuts along the high-symmetry directions and constant $\textbf{Q}$ cuts at the high-symmetry points.} \textbf{a,} Constant energy cuts along the $\Gamma$-M-K-$\Gamma$ direction at the indicated energy at 0.05 K. The solid curves indicate the simulated spin excitations using the model specified in the text and the vertical dashed lines indicate the high-symmetry points. \textbf{b-d,} Constant $\textbf{Q}$ cuts at the K, M and $\Gamma$ points at various temperatures. The data presented in \textbf{a}, \textbf{c} and \textbf{d} were measured with ${E_\textrm{i} = 4.8}$ meV while the data in \textbf{b} were measured with ${E_\textrm{i} = 1.7}$ meV. The gray dashed lines indicate the elastic incoherent scattering. The error bars indicate the standard deviation.
}
\label{fig4}
\end{figure}

The observation of a magnetic Bragg peak at the K point instead of the M point is in clear contradiction with the stripe ordered ground state for the Ising interactions~\cite{TMGO_YL}. To determine the nature of the K point order, we measured the spin excitations over a wide range of momentum in TmMgGaO$_4$. Fig.~\ref{fig2} shows the momentum dependence of the spin excitations at various energies. The constant energy images at low energies display sharp spin response at the K points (Fig.~\ref{fig2}a). As the energy increases, the spin excitation disperses outward from the K points and then forms ring-like patterns around the $\Gamma$ points at higher energies (Fig.~\ref{fig2}b-e). Eventually, the spectra reach the band top at the $\Gamma$ points and vanish above $1.7$ meV (Fig.~\ref{fig2}f). The overall dispersion can be seen more clearly in the energy dependence of the spectral intensity along the high-symmetry directions in Fig.~\ref{fig3}a, which reveals a very sharp spin-wave spectrum. This suggests that the exchange disorder induced by the Mg/Ga site mixing in the non-magnetic layers is not significant, in stark contrast to the conjecture that the Mg/Ga site mixing will introduce strong exchange variation (ref.~\onlinecite{Wen}). The spin excitation dispersion is further confirmed by a series of constant-energy cuts through the high-symmetry points (Fig.~\ref{fig4}a) and constant-\textbf{Q} cuts (Fig.~\ref{fig4}b). In addition, the constant-\textbf{Q} cut at the K point reveals a spin-gap feature at low energies (Fig.~\ref{fig4}b). This gap is gradually closed with increasing temperature (Fig.~\ref{fig4}b). Meanwhile, the spectral weight is transferred to lower energies, turning into quasi-elastic excitations at higher temperatures (Fig.~\ref{fig4}b-d).

\textbf{Discussion}

To describe the momentum and energy dependence of the spin excitation spectra, we adopt the linear spin wave theory (LSW). For conventional magnetic systems in which all $S^{\mu}_i$ components exhibit dipolar properties, neutron scattering will probe all three spin channels, $S_{xx}$, $S_{yy}$ and $S_{zz}$, since the neutron is scattered by the magnetic moment through dipole-dipole interactions. Correspondingly, we calculate the spin wave excitations using the \textsc{spinw} program \cite{spinw} for the pure Ising model, XY model and Heisenberg model (Supplementary Note 2). However, none of these models with a dipolar order is consistent with our data. For example, for the XY model that supports the 120 degree N\'eel order with magnetic Bragg peaks at K points, the spin wave dispersion around $\Gamma$ points should exhibit a minimum instead of a maximum that is observed in our experiments. In addition, the Ising model with a stripe-type order that has been proposed in ref.~\onlinecite{TMGO_YL}, however, will lead to magnetic Bragg peaks at M points and dispersionless spin excitations, clearly inconsistent with our data.

On the other hand, an intertwined multipolar and dipolar order is in excellent agreement with the experimental data. In TmMgGaO$_{4}$, the large $g$-factor in the magnetization experiments indicates strong Ising character of the Tm$^{3+}$ ions, suggesting that the wave functions are dominated by $J^{z}=\pm$6 components \cite{TMGO_YL}. Under the $D_{3d}$ symmetric CEF, these $J^{z}=\pm$6 dominated components are not allowed to form a symmetry protected doublet. Instead, they must hybridize into two non-magnetic singlets and open an energy gap. The full singlet wave functions dictating the $D_{3d}$ CEF symmetry must take the following form:

\begin{align}
{\ket{\Psi_{i}^{+}}}\sim & c_{6}\left({\ket{+6}}_{i}+{\ket{-6}}_{i}\right)+c_{3}\left({\ket{+3}}_{i}-{\ket{-3}}_{i}\right)+c_{0}{\ket{0}}_{i},\label{eq:psiu}\\
{\ket{\Psi_{i}^{-}}}\sim & c_{6}'\left({\ket{+6}}_{i}-{\ket{-6}}_{i}\right)+c_{3}'\left({\ket{+3}}_{i}+{\ket{-3}}_{i}\right)\label{eq:psid}
\end{align}

Here, these two singlets, ${\ket{\Psi_{i}^{+}}}$ and ${\ket{\Psi_{i}^{-}}}$, carry $A_{1g}$ and $A_{2g}$ representation of the $D_{3d}$ group, respectively. $\ket{j_{z}}$ denotes
for $J^{z}=j_{z}$ states at site $i$ and $c_{6}$, $c_{3}$, $c_{0}$, $c_{6}'$ and $c_{3}'$ are real numbers with $|c_{6}|\approx|c_{6}'|\gg c_{3},c_{3}',c_{0}$. For simplicity, in the following the site index $i$ will be omitted. The local moment can be constructed by the following spin-1/2 operators acting on the quasi-doublet:

\begin{align}
S^x&=\frac{i}{2}(\ket{\Psi^-}\bra{\Psi^+}-\ket{\Psi^+}\bra{\Psi^-}),\\
S^y&=\frac{1}{2}(\ket{\Psi^+}\bra{\Psi^+}-\ket{\Psi^-}\bra{\Psi^-}),\\
S^z&=\frac{1}{2}(\ket{\Psi^+}\bra{\Psi^-}+\ket{\Psi^-}\bra{\Psi^+})
\end{align}

where our definition of the spin operators here is a bit different from the conventional choice. Our choice is designed for the particular bases and the wave functions of two singlet states in TmMgGaO$_4$. The transformation of the symmetry operators under CEF symmetry operations (three-fold rotation about $c$ axis $C_{3}$ and time reversal $T$) take the following form:

\begin{align}
C_{3}: & S^{x}\rightarrow S^{x},S^{y}\rightarrow S^{y},S^{z}\rightarrow S^{z}\\
T: & S^{x}\rightarrow S^{x},S^{y}\rightarrow S^{y},S^{z}\rightarrow-S^{z}
\end{align}

We can see that the transverse components $S^{x}$ and $S^{y}$ are even under time-reversal and stay invariant under the three-fold rotation. These unique properties indicate that the transverse components are multipoles with rank $\ge3$ that are not directly accessible for neutrons, nor other conventional experimental probes. Meanwhile, the longitudinal component $S^{z}$ is related to dominated $\Delta J_{z}=0$ process (e.g., from $\ket{+6}$ to $\ket{+6}$ or from $\ket{-6}$ to $\ket{-6}$). Therefore, it behaves as a conventional magnetic dipole along the $c$ axis, and can directly couple to external magnetic field or neutron spins.

To keep the essential physics of this system it is enough to consider only the Ising-type spin interactions (since the local moments have strong Ising nature) and the CEF splitting between singlets here. With the gap between singlets captured by a transverse field term along $y$ axis ($-h\sum_{i}S_{i}^{y}$) and the Ising interactions taken into account up to second nearest-neighbour, the effective Hamiltonian reads:

\begin{align}
\mathcal{H} &= \sum_{\langle ij \rangle}J^{zz}_1S^z_i S^z_j+\sum_{\langle\langle ij \rangle\rangle}J^{zz}_2S^z_i S^z_j-h\sum_{i}S^y_i,
\end{align}
where $\langle ij \rangle$ and $\langle\langle ij \rangle\rangle$ denote the nearest and next-nearest neighbours, respectively (Supplementary Fig. 2a). Despite the strong Ising nature, the CEF splitting term can create strong quantum fluctuations upon the spins. Based on this model, we re-calculate the spin wave dispersion and find a set of parameters that can accurately describe our data: $J^{zz}_1$=0.54(2) meV, $J^{zz}_2$=0.026(6) meV, $h$=0.62(2) meV. The calculated \textit{E}-\textbf{k} relationship shows excellent agreement with the experimental observation (Fig.~\ref{fig2}g-l, \ref{fig3}b). The corresponding magnetic structure is a three-sublattice structure in which the hidden components form a preformed ferro-multipolar order along the $y$ direction in the effective spin space due to the polarization effect from the transverse field while the out-of-plane dipole moments order antiferromagnetically at K points (Fig.~\ref{fig1}j). The out-of-plane dipolar order (spin-up, spin-down, spin-0 in the three sublattices, respectively) shows no macroscopic magnetization and is indicated by our neutron diffraction measurements that reveal a magnetic Bragg peak at the K point. The multipoles, however, do not linearly couple to neutrons and therefore are hidden in the neutron diffraction measurements.

Although these multipolar components are not directly visible, the elementary excitations of the multipolar components can be accessible in inelastic neutron experiments. As the multipolar components do not commute with the dipolar ones, measuring the $S^z$ moment will induce spin-flipping events on the multipolar components, leading to coherent multipolar spin wave excitations~\cite{GCoctu,GCnonK}. As a result, the LSW theory remains a valid description for the excitations but only the longitudinal $S_{zz}$ channel is involved.

Because multipoles have a complex structure with no spatially uniform magnetization, the multipolar spin wave may display an anisotropic and non-monotonic \textbf{Q}-dependence of the magnetic form factor which is different from a conventional dipole spin wave~\cite{MultiRev,CeLaBNS1}. Indeed, Fig. 2 shows that the spin excitations in the second Brillouin zone are not weaker than those of the first Brillouin zone, in stark contrast to a conventional dipolar form factor of Tm$^{3+}$ that decreases with increasing \textbf{Q}. This further confirms the multipolar nature of the spin excitations (Supplementary Note 3). In addition, because the multipolar moments do not couple directly to the external field, the in-plane magnetization should be invisible to the conventional magnetic probe in the intertwined dipolar and ferro-multipolar order state in TmMgGaO$_4$. This is indeed the case as is illustrated in Fig.~\ref{fig1}a,~\ref{fig1}b: little magnetic susceptibility is observed when the external magnetic field is applied in the $ab$ plane. We note that the residual tiny in-plane signal observed could be either due to the imperfection of the crystal alignment for the measurements or due to a magnetic response from higher order Van-Vleck like process.

More insight into the nature of the hidden order and local moment structures of f-electron materials can be obtained by comparing TmMgGaO$_4$ with other hidden order materials. In URu$_2$Si$_2$, neutron scattering experiments suggest that both the commensurate and incommensurate spin excitations only display longitudinal components~\cite{URuSi_polarized}. This is in analogy to our data where the transverse excitations are absent due to the multipolar behavior. Moreover, although the hidden order parameters are in general invisible to standard neutron diffraction experiments, weak Bragg peaks may appear in certain materials presumably owing to the higher-order interactions between neutron spins and the hidden order parameter, which is the case for Ce$_{1-x}$La$_x$B$_6$ and URu$_2$Si$_2$ (ref.~\onlinecite{URuSiNS1,CeB_polarized}). We therefore carefully performed neutron diffraction survey near the ferro-multipolar order wavevector $\Gamma$ in TmMgGaO$_4$. Indeed, a very weak Bragg peak at $\Gamma$ is revealed at 0.485 K. The peak is about three orders of magnitude weaker than the magnetic Bragg peak at the K point, implying that it is indeed possible due to a higher-order interaction (Fig.~\ref{fig1}f). The peak weakens with increasing temperature in a similar manner to the magnetic Bragg peak at the K point (Fig.~\ref{fig1}g). This is consistent with the presence of an intertwined ferro-multipolar and dipolar order in this compound.

The similar properties between insulating TmMgGaO$_4$ and metallic heavy-fermion hidden-order materials could have important implications. The multipolar behavior in f-electron systems mainly arises from the orbital degrees of freedom in the presence of strong SOC and CEF. In insulating TmMgGaO$_4$, the degeneracy of the ground state is lifted by the CEF effect, leading to a preformed ferro-multipolar order. On the other hand, in the case of heavy fermion systems such as CeB$_6$, the degeneracy of the ground state quartet is likely lifted by the antiferromagnetic interactions between the localized 4$f$ electrons that are possibly mediated by the itinerant conduction electrons. In this sense, the multipolar order can be viewed as an orbital order due to the lifting of the degenerate ground states with or without itinerant electrons. This implies that similar approach and analysis used here could potentially apply in other heavy-fermion hidden-order materials.

\textbf{Methods}

\textbf{Sample synthesis.} Polycrystalline TmMgGaO$_4$ samples were synthesized through the solid-state-reaction method. The starting materials Tm$_2$O$_3$, MgO and Ga$_2$O$_3$ were mixed in stoichiometric quantities and heated in ceramic crucibles at 1500 $^\circ$C for 4 days in air. The polycrystalline samples were then loaded in rubber tubes and compressed into rods hydrostatically at 300 MPa, which were further sintered at 1500 $^\circ$C for 8 hours. The resulting feed rods were then transferred into an optical floating zone furnace for the crystal growth. Counter-rotation of feed rod (24 rpm) and crystal (29 rpm) has been
applied during the growth. Large and high quality TmMgGaO$_4$ single crystals up to 5 cm in length can be obtained with a growth rate of 0.6 mm/h (Supplementary Note 1).

\textbf{Neutron scattering experiments.} The neutron scattering experiments were carried out on the cold three axes spectrometer PANDA at the Heinz Maier-Leibnitz Zentrum, Garching, Germany, and cold neutron multi-chopper spectrometer LET at the Rutherford Appleton Laboratory, Didcot, UK. For the PANDA experiment, we used a vertical focused PG(002) as a monochromator and analyser; the final neutron energy was fixed at $E_\textrm{f}$=4.06 meV, resulting in an energy resolution of around 0.1 meV. A Be filter is placed after the sample to avoid the contamination from neutrons with higher orders. One piece of single crystal (3 g) was aligned in the ($HK0$) plane for the measurements. A closed-cycle refrigerator equipped with a $^3$He insert was used to reach the base temperature of 485 mK. For the LET time-of-flight neutron scattering experiment, we chose the incident energies of 11.0, 4.8, 2.7 and 1.7 meV with energy resolutions of 0.343, 0.114, 0.047 and 0.027 meV, respectively. A dilution insert for the standard $^4$He cryostat was used to reach the base temperature of 50 mK. Six pieces of single crystals with a total mass of 17.2 g were co-aligned in the ($HK0$) plane for the measurements. The data were analysed using the Horace-Matlab suite \cite{Horace}.

All neutron data are presented without background subtraction or symmetrizing. With background subtraction or symmetrizing, the data do not show significant change.

\newpage

\newpage
\renewcommand\figurename{\textbf{Supplementary Figure}}
\setcounter{figure}{0}
\setcounter{page}{1}

\begin{center}

Supplementary Information for

\textbf{Intertwined dipolar and multipolar order in the triangular-lattice magnet TmMgGaO$_4$}

Shen \textit{et al.}

\end{center}

\textbf{Supplementary Note 1: Sample synthesis and characterization.}

High quality single-crystalline TmMgGaO$_4$ samples were synthesized using the floating-zone technique. The single crystal is transparent with shiny cleaved surfaces (Supplementary Fig. 1a). Sharp and clear diffraction spots can be seen in the X-ray Laue pattern (Supplementary Fig. 1b). In the single-crystal X-ray diffraction measurement, a series of reflections along the $L$ direction can be indexed and the full-width at the half maximum (FWHM) of the rocking scan across \textbf{Q}=(0, 0, 18) is around 0.025 degrees (Supplementary Fig. 1c). These results indicate the high crystallization quality of the single crystals.

\textbf{Supplementary Note 2: Calculations of the spin excitations in TmMgGaO$_4$.}

In order to understand the observed spin excitation spectra in TmMgGaO$_4$ at low temperature, we calculate the spin wave dispersion by linear spin wave theory (LSW) using the \textsc{spinw} program \cite{spinw}. Here we present the calculated results of five representative scenarios.

First, we examine the possibilities of conventional dipolar ordering. The corresponding spin Hamiltonian is assumed as:
\begin{align}
\mathcal{H} &= \sum_{\langle ij \rangle}[J^{zz}_1S^z_i S^z_j
+J^\pm_1(S^+_i S^-_j+S^-_i S^+_j)]
+\sum_{\langle\langle ij \rangle\rangle}J^{zz}_2S^z_i S^z_j
\end{align}
where $S^{\pm}=S^x\pm iS^y$ and $J_1$ and $J_2$ are nearest and next-nearest neighboured magnetic interactions, respectively, as illustrated in Supplementary Fig. 2a. The three-fold magnetic twins are included for all calculations.

We consider the isotropic Heisenberg model ($J^{zz}_1$ = 2$J^{\pm}_1$ = 0.9 meV, $J^{zz}_2$ = 0 meV) and anisotropic XY model ($J^{zz}_1$ = $J^{zz}_2$ = 0 meV, $J^{\pm}_1$ = 0.4 meV) that support the 120 degree N\'eel order and lead to strong magnetic Bragg peaks at K points (Supplementary Fig. 3a). We calculate both the total scattering function ($S_{xx} + S_{yy} + S_{zz}$) and the longitudinal component ($S_{zz}$) for both of the models (Supplementary Fig. 4a-d). Although the calculated spin waves catch the feature of Goldstone mode stemmed from K points, the branches around $\Gamma$ points go as a minimum in the calculation instead of a maximum which is observed in the neutron experiments. Moreover, the branches around the M points are located at higher energies than the experimental data.

Another scenario is the Ising model with a stripe order that has been proposed in a recent research\cite{TMGO_YSLi} (Supplementary Fig. 3b). The parameters are chosen to be $J^{zz}_1$ = 0.8 meV and $J^{zz}_2$ = 0.076 meV. The proposed stripe antiferromagnetic structure will lead to magnetic Bragg peaks at M points. We measured the elastic scattering at M points with the similar statistic with that at $\Gamma$ point in the main text. As shown in Supplementary Fig. 5, no indication of peak-like feature can be distinguished. For the spin excitations, since all spins are aligned along the $c$ direction, the $S_{zz}$ sector vanishes for local spins and the transverse components are essentially dispersionless, clearly inconsistent with our data (Supplementary Fig. 4e).

In addition, we calculate the energy dependence of the spectral intensity at $\Gamma$ and M points with various values of $J_1^{zz}/J_1^{\pm}$. It is shown that the simulated spin wave energy at the $\Gamma$ point is either comparable or lower than that at the M point (Supplementary Fig. 6). This is inconsistent with our data where the spin wave energy at the $\Gamma$ point exhibits a maximum with an energy significantly higher than that at the M point (Fig. 3a). Thus, the anisotropic XXZ model with a dipolar order by no means can describe the spin wave dispersion in TmMgGaO$_4$.

We also calculate the scattering function based on the quasi-doublet scenario that is proposed in the main text (Supplementary Fig. 2b). We show that the longitudinal excitations ($S_{zz}$ channel) of the intertwined multipolar order are in excellent agreement with our data (Fig. 2g-l, Fig. 3b).

Finally, we consider the intertwined multipolar order raised from a non-Kramers doublet system \cite{GCNonK}. It is described by the Hamiltonian
\begin{align}
\mathcal{H} &= \sum_{\langle ij \rangle}[J^{zz}_1S^z_i S^z_j+J^\pm_1(S^+_i S^-_j+S^-_i S^+_j)+J^{\pm \pm}_1(\gamma_{ij}S^+_i S^+_j+\gamma^*_{ij}S^-_i S^-_j)]+\sum_{\langle\langle ij \rangle\rangle}J^{zz}_2S^z_i S^z_j
\end{align}
where $\gamma_{ij}$ are phase factors that depend on the directions of the bonds. In this scenario, the effective spin components $S^x$ and $S^y$ show hidden quadrupolar behavior while the out-of-plane component $S^z$ shows dipolar behavior, similar to the quasi-doublet scenario that is proposed for TmMgGaO$_4$ in the main text. However, the observed large Land\'e $g$-factor (12.11) is inconsistent with a non-Kramers doublet state. Moreover, the heat capacity of the highly diluted Tm$_{0.04}$Lu$_{0.96}$MgGaO$_4$ shows a finite zero-temperature limit of $C_m/T$, which is consistent with the quasi-doublet state, but inconsistent with the non-Kramers doublet state \cite{TMGO_YSLi}.

\textbf{Supplementary Note 3: Magnetic form factor.}

For conventional dipolar moments, the magnetic form factor decreases with increasing \textbf{Q} while the multipoles may display an anisotropic and non-monotonic \textbf{Q}-dependence of the magnetic form factor, because multipoles have complex magnetization density \cite{MultiRev}. In Supplementary Fig. 7, we present the constant energy slices of the simulated spin wave excitations considering a dipolar Tm$^{3+}$ form factor. The simulated spin excitation signal in the second Brillouin zone is considerably weaker than that of the first zone, which is clearly inconsistent with the measured data in Fig. 2a-f. The constant energy cuts through the first and second Brillouin zones further prove that the spin excitation intensity does not follow the dipolar Tm$^{3+}$ form factor (Supplementary Fig. 8).

Correspondence and requests for materials should be
addressed to J.Z. (zhaoj@fudan.edu.cn) or G.C. (gchen\_physics@fudan.edu.cn).

\newpage

\begin{figure}[htbp]
\includegraphics[width=0.7\textwidth]{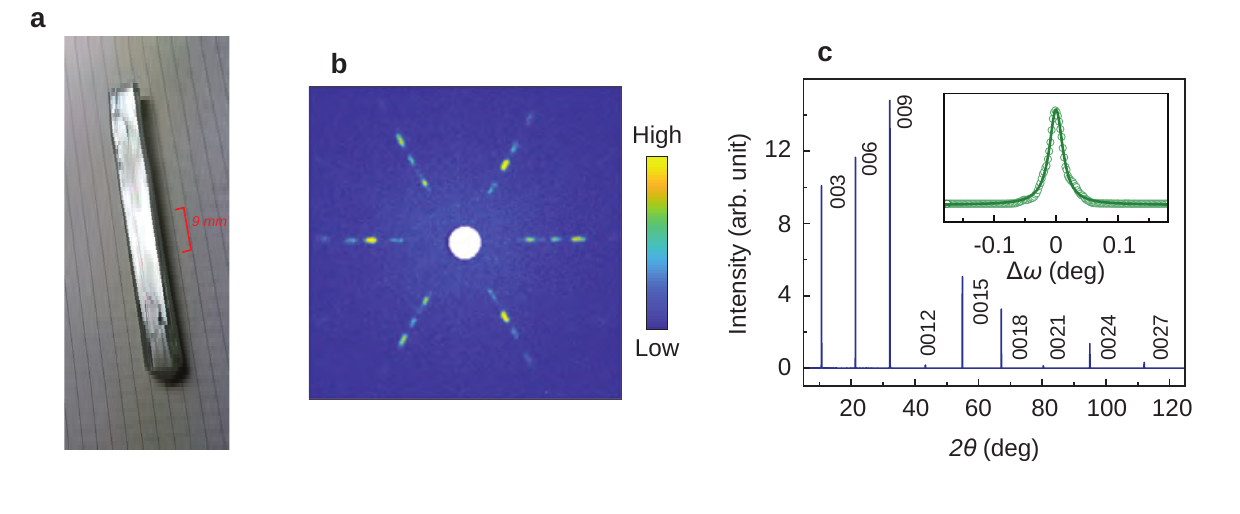}
\caption{\textbf{Photograph and X-ray diffraction patterns of TmMgGaO$_4$ single crystals.} \textbf{a,} A photograph of a representative TmMgGaO$_4$ single crystal. \textbf{b,} X-ray Laue pattern viewed from the $c$ axis. The color bar indicates scattering intensity in arbitrary unit in linear scale. \textbf{c,} X-ray diffraction pattern from the cleaved surface of a TmMgGaO$_4$ single crystal. The inset shows the Lorentz fitting of the rocking curve of the (0, 0, 18) peak.
}
\end{figure}

\begin{figure}[htbp]
\includegraphics[width=0.6\textwidth]{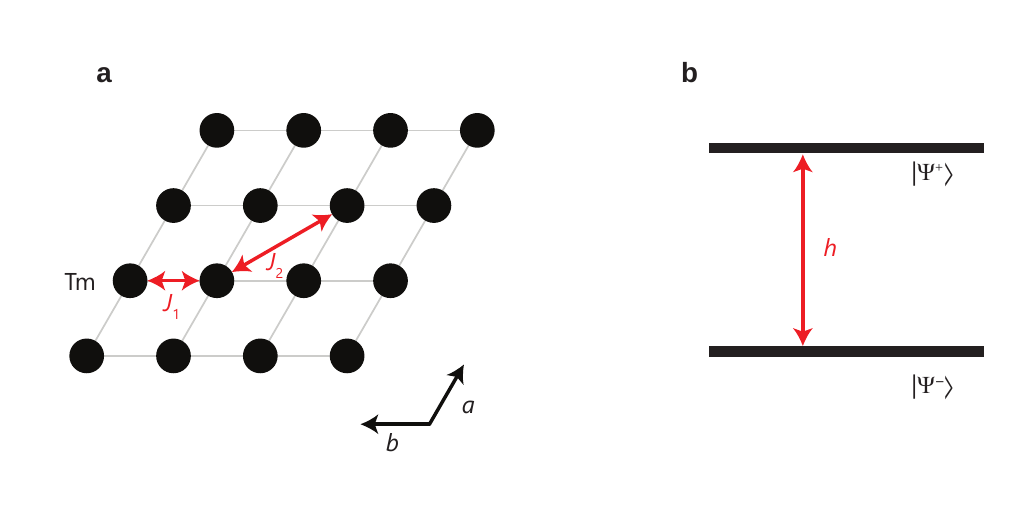}
\caption{\textbf{Sketch of the magnetic interactions and CEF splitting in TmMgGaO$_4$.} \textbf{a,} Illustration of the triangular lattice plane of Tm ions. The nearest ($J_1$) and next-nearest magnetic interactions ($J_2$) are marked by the red arrows. The grey solid lines indicate the unit cell. \textbf{b,} CEF splitting of the low-lying two singlet levels. $h$ is the effective transverse field term that models the splitting between two singlet levels.
}
\end{figure}

\begin{figure}[htbp]
\includegraphics[width=0.8\textwidth]{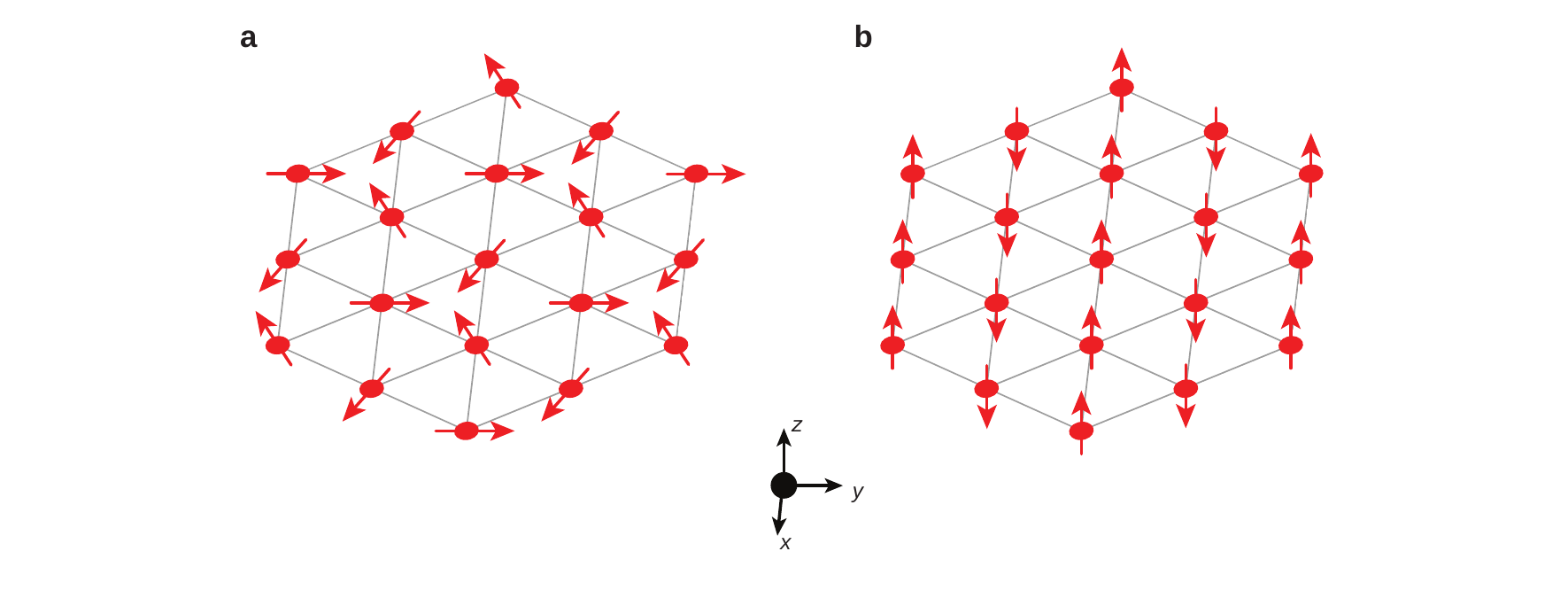}
\caption{\textbf{Mean-field spin configurations of the different states.} \textbf{a,} Illustration of the spin structure in 120 degree N\'eel order. All the spins lie in the $xy$ plane. \textbf{b,} Magnetic structure of the stripe order. All the spins are aligned along $z$ direction.
}
\end{figure}

\begin{figure}[htbp]
\includegraphics[width=1\textwidth]{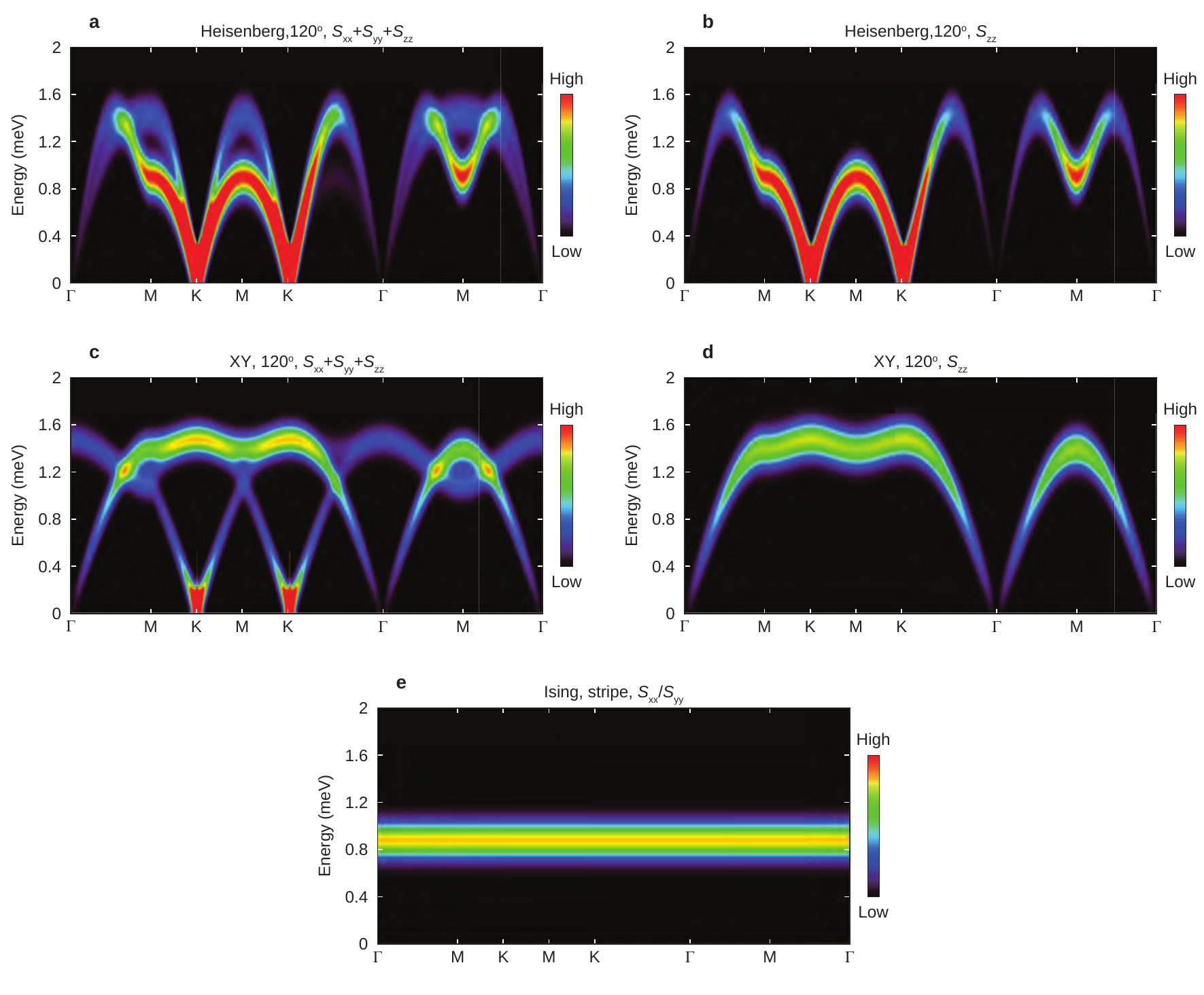}
\caption{\textbf{Calculated spin excitation dispersions with different models and scattering sectors.} \textbf{a, b,} Total and longitudinal scattering function for isotropic Heisenberg model. \textbf{c, d,} Total and longitudinal scattering function for XY model with the 120 degree magnetic structure. \textbf{e,} Transverse scattering sector for Ising model. The color bars indicate scattering intensity in arbitrary unit in linear scale.
}
\end{figure}

\begin{figure}[htbp]
\includegraphics[width=0.6\textwidth]{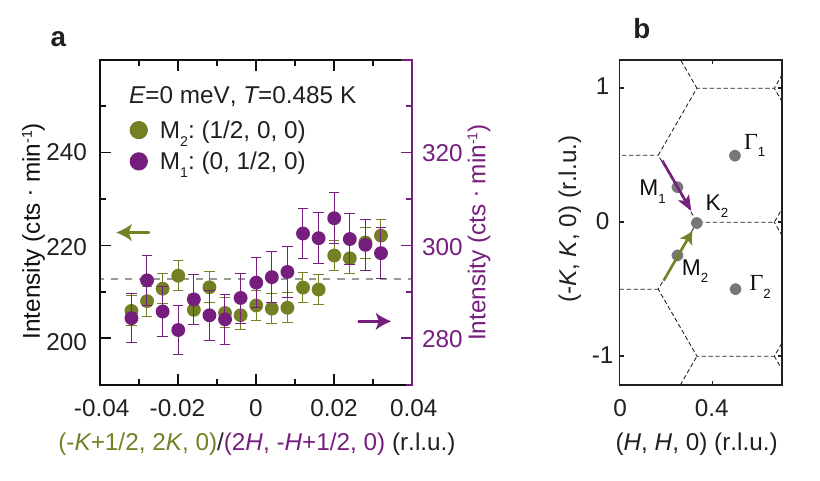}
\caption{\textbf{Absence of Bragg peaks at M points.} \textbf{a,} Constant energy cuts across M points at $E$=0 meV along the directions indicated by the yellow and purple arrows in \textbf{b}. Error bars, 1 s.d. \textbf{b}, Sketch of the reciprocal space.
}
\end{figure}

\begin{figure}[htbp]
\includegraphics[width=0.7\textwidth]{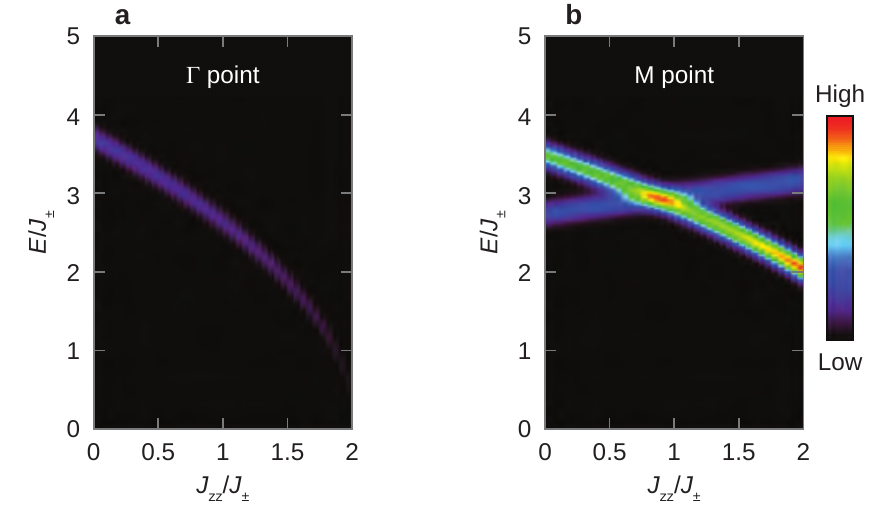}
\caption{\textbf{Calculated spin wave excitations at $\Gamma$ and M points.} \textbf{a, b,} Energy dependence of the spectral intensity at $\Gamma$ and M points with various magnetic interactions. The color bar indicates scattering intensity in arbitrary unit in linear scale.
}
\end{figure}

\begin{figure}[htbp]
\includegraphics[width=1\textwidth]{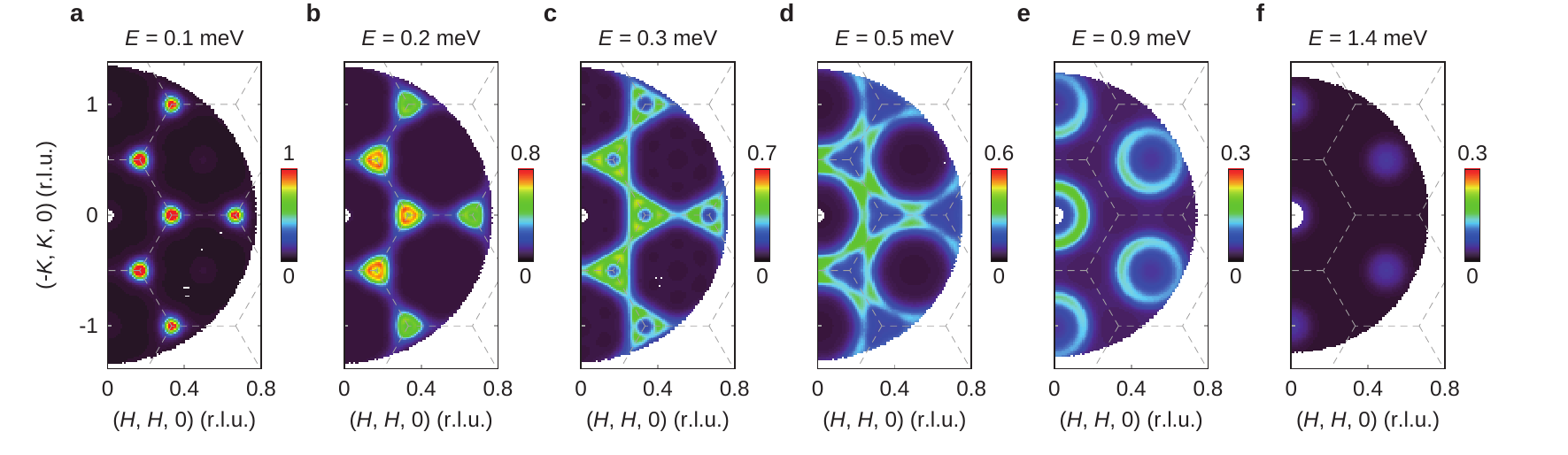}
\caption{\textbf{Calculated momentum dependence of spin excitations at the indicated energies assuming a dipolar Tm$^{3+}$ form factor.} \textbf{a-f,} Calculated spin excitations using the model and parameters specified in the text. The color bars indicate scattering intensity in arbitrary unit in linear scale.
}
\end{figure}

\begin{figure}[htbp]
\includegraphics[width=0.6\textwidth]{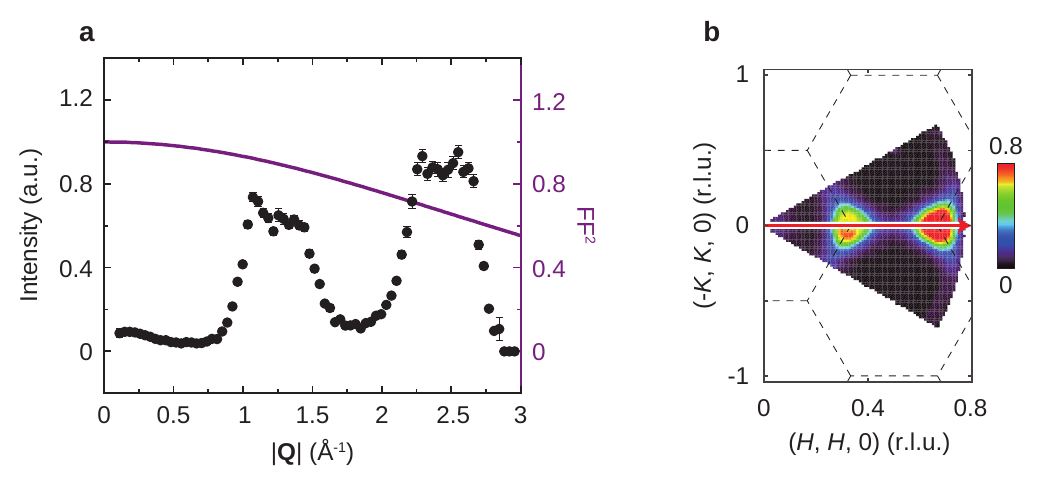}
\caption{\textbf{$Q$-dependence of the spin wave intensity.} \textbf{a,} Constant energy cuts at 0.05 K and 0.2 meV along the direction marked by the red arrow in \textbf{b}. The purple solid lines indicate the form factor of dipolar Tm$^{3+}$ ions. \textbf{b,} Constant energy slice at 0.05 K and 0.2 meV. All the data are symmetrized into a 60 degree wedge of the reciprocal lattice. The color bar indicates scattering intensity in arbitrary unit in linear scale.
}
\end{figure}

\end{document}